\documentclass[12pt]{article}
\usepackage{graphicx}
\usepackage{array}
\usepackage{multirow}
\usepackage{amsfonts}
\usepackage{amssymb}
\usepackage{mathrsfs}
\usepackage{amsmath}
\usepackage{subfigure}
\UseRawInputEncoding
\makeatletter

\newcommand{\Rmnum}[1]{\expandafter\@slowromancap\romannumeral #1@}
\makeatother

\usepackage[unicode]{hyperref}
\usepackage[numbers,sort&compress]{natbib}

\begin{document}
\renewcommand{\thefootnote}{\fnsymbol{footnote}}
\begin{titlepage}

\vspace{10mm}
\begin{center}
{\Large\bf Black holes in Einstein-Gauss-Bonnet gravity with a background of modified Chaplygin gas}
\vspace{9mm}

{{\large Xiang-Qian Li${}^{1,}$\footnote{E-mail address: lixiangqian@tyut.edu.cn}},
{\large Bo Chen${}^{1,}$\footnote{E-mail address: chenbo@tyut.edu.cn}},
{\large Li-Li Xing${}^{1,}$\footnote{E-mail address: xinglili@tyut.edu.cn}}\\

\vspace{6mm}
${}^{1}${\normalsize \em College of Physics, Taiyuan University of Technology, Taiyuan 030024, China
}
}

\end{center}

\vspace{10mm}
\centerline{{\bf{Abstract}}}
\vspace{6mm}
\noindent

Supposing the existence of modified Chaplygin gas with the equation of state $p=A\rho-B/\rho^\beta$ as a cosmic background, we obtain a static spherically-symmetric solution to the Einstein-Gauss-Bonnet gravitational equations in 5D spacetime. The spacetime structure of the obtained black hole solution could be asymptotically anti-de Sitter or de Sitter, according to the specific values of modified Chaplygin gas parameters versus the cosmological constant. We analyze the parametric regions for both kinds of solutions. For asymptotically anti-de Sitter black hole, there exists the so-called small/large black hole phase transition, we obtain critical values of pressure, volume, and temperature and investigate the effects of both the Gauss-Bonnet gravity and the modified Chaplygin gas on these values. For asymptotically de Sitter black hole, no $P-r_h$ criticality and phase transition appear, and we show that the thermodynamic systems related to various horizons of asymptotically de Sitter black hole are in fact entangled.

\vskip 20pt
\noindent
{\bf PACS Number(s)}: 04.20.Cv, 04.50.Gh, 95.35.+d


\end{titlepage}
\newpage
\renewcommand{\thefootnote}{\arabic{footnote}}
\setcounter{footnote}{0}
\setcounter{page}{2}

\section{Introduction}

Ones of the natural modifications of general relativity (GR) in higher dimensions are the Lovelock theories where higher-curvature terms are supplemented \cite{Lovelock}. Taking into account the first additional term of general Lovelock theory, i.e. the Gauss-Bonnet invariant, to Einstein gravity, one obtains the Einstein-Gauss-Bonnet (EGB) gravity. It's interesting that EGB gravity can also be arisen from the low-energy limit of heterotic string theory \cite{freeghost,GrossWittennpb1986,Metsaevplb1987}. It is believed that EGB gravity can avoid some of the shortcomings of Einstein gravity \cite{StelleGRG1978,MalufGRG1987,FarhoudiGRG2006}. In addition, the EGB gravity consists of the Einstein-Hilbert action plus curvature-squared terms (see Refs. \cite{BoulwarePRL1985,ZuminoPR1986,MyersNPB1987,MyersNPB1989,NeupaneNPB2002,RGCaiPRD2002} for more details), leading to field equations with no more than second derivatives of the metric, thus free of ghost \cite{freeghost}. Considering the EGB gravity context, black hole solutions and their thermodynamical behaviors have been investigated in much literature.
The spherically symmetric black hole solutions in EGB gravity have been found in \cite{BoulwarePRL1985,WheelerNPB1986}. The EGB black holes in anti de Sitter and de Sitter spaces have been discussed separately in \cite{ChoPRD2002} and \cite{RGCaiPRD2004}. Ref.~\cite{DottiPRD2007} presented an exhaustive classification of static solutions for the five-dimensional EGB theory in vacuum. Recently, Refs.~\cite{Doneva:2017bvd,Silva:2017uqg,Blazquez-Salcedo:2018jnn,Macedo:2019sem,Doneva:2019vuh} concentrated on black hole solutions in 4D Gauss-Bonnet-scalar gravity with coupling between the scalar and the Gauss-Bonnet invariant.

Matter content of the Universe is still an unsolved problem in the framework of modern cosmology. The latest release of 2018 Planck full-sky maps about the CMB anisotropies \cite{Planck2018} illustrates that baryon matter component is no more than $5\%$ for total energy density. In contrast, the invisible dark components, including dark energy and dark matter, are about $95\%$ energy density in the Universe. The dominance of the dark sector over the Universe makes the study of black holes surrounded by these mysterious field well-deserved. Quintessence is a possible candidate for dark energy, which is characterized by the linear equation of state $p_q=w\rho_q$. Significant attention has been devoted to discussion of static spherically-symmetric black hole solutions surrounded by quintessence matter and their properties \cite{KiselevCQG2003,MaCPL2007,FernandoGRG2012,FengPLA2014,MalakolkalamiASS2015,HussainGRG2015,Ghosh:2016ddh,Ghosh:2017cuq,Toledo:2018pfy}, within which, Ref. \cite{Ghosh:2016ddh} paid attention to the context of EGB gravity. Except the quintessence matter, many authors have found exact black hole solutions in EGB gravity with some other sources. Ref.~\cite{DominguezPRD2006} showed a class of dynamical black hole solutions in EGB gravity by restricting the energy-momentum tensor with some constraints. Ref.~\cite{WiltshirePRD1988} derived spherically-symmetric solutions in EGB gravity with a Born-Infeld term. Ref.~\cite{MaedaPRD2009} derived electrically charged EGB black hole solutions with a nonlinear electrodynamics source given as an arbitrary power of the Maxwell invariant, and Ref.~\cite{HendiEPJC2014} obtained the topological black hole solutions in the presence of another two classes of nonlinear electrodynamics source. Ref.~\cite{MazharimousaviPRD2007} represented EGB black holes with a background of Yang-Mills fields. Ref.~\cite{HerscovicPLB2010} obtained a black hole solution of the 5D EGB theory for the string cloud model.

With regard to the Universal dark sector, there exists another possibility that the unknown energy component is a unified dark fluid which mixes dark matter and dark energy. Among the proposed unified dark fluid models, the Chaplygin gas \cite{Kamenshchik:2001cp} and its generalized model \cite{Bilic:2001cg,Bento:2002ps} have been widely studied in order to explain the accelerating Universe \cite{Carturan:2002si,Amendola:2003bz,Bean:2003ae}. It is worth noting that the equation of state of the original Chaplygin gas model can be obtained from the Nambu-Goto action for a $D$-brane moving in a ($D+2$)-dimensional spacetime in light-cone parametrization~\cite{Ogawa:2000gj}. Ref.~\cite{Lovelockdf2019} considered a model that charged static spherically-symmetric black hole is surrounded by Chaplygin-like dark fluid in the framework of Lovelock gravity. In this paper we study the static spherically-symmetric black holes surrounded by the modified Chaplygin gas (MCG) with the equation of state $p=A\rho-B/\rho^\beta$ in the 5-dimensional (5D) EGB gravity.

The plan of this paper is as follows. In section \ref{section2}, for the MCG in 5D spacetime, we deduce its energy momentum tensor, with the help of which we obtain the static spherically-symmetric solutions to the EGB gravitational equations. We also study the effects of MCG parameters on black hole solutions and analyze the parametric regions for asymptotically (anti-)de Sitter black hole. Further we study the thermodynamical properties of the obtained black hole solutions in section \ref{section3}. Section \ref{section4} gives the conclusion. For completeness, we give the MCG-surrounded-EGB black hole solution and related thermodynamical quantities in \emph{D}-dimensional spacetime with Appendix \ref{appendixA}.

We use units which fix the speed of light and the 5D gravitational constant via $8\pi G_5 = c = 1$, and use the metric signature ($-,\;+,\;+,\; \cdots,\;+$).

\section{Surrounded black hole solutions in Einstein-Gauss-Bonnet gravity}
\label{section2}

\subsection{The Einstein-Gauss-Bonnet theory}
The Lovelock theory is an extension of the general relativity to higher-dimensions. In this theory the first and second order terms correspond to the Einstein-Hilbert and Gauss-Bonnet terms, respectively.
The action for 5D EGB gravity with matter field reads
\begin{equation}\label{action}
\mathcal{S}=\frac{1}{2}\int_{\mathcal{M}}d^{5}x\sqrt{-g}\left[  \mathcal{L}_{E} +\alpha \mathcal{L}_{GB}
 -2\Lambda \right] + \mathcal{S}_{M}.
\end{equation}
$\mathcal{S}_{M}$ denotes the action associated with matter and $\alpha$ is coupling constant that we assume to be non-negative. The Einstein term is $ \mathcal{L}_E = R$, and the second-order Gauss-Bonnet term $\mathcal{L}_{GB}$ is
\begin{equation}
\mathcal{L}_{GB}=R_{\mu\nu\gamma\delta}R^{\mu \nu\gamma\delta}-4R_{\mu\nu}R^{\mu\nu}+R^{2}.
\end{equation}
Here, $R_{\mu\nu}$, $R_{\mu\nu\gamma\delta\text{}}$, and $R$  are the Ricci tensor, Riemann tensor, and  Ricci scalar, respectively. The variation of the action with respect to the metric $g_{\mu\nu}$ gives the EGB equation:
\begin{equation}\label{ee}
G_{\mu\nu}\equiv G_{\mu\nu}^{E}+\alpha G_{\mu\nu}^{GB}+\Lambda g_{\mu\nu}=T_{\mu\nu},
\end{equation}
where $G_{\mu\nu}^{E}$ is the Einstein tensor while $G_{\mu\nu}^{GB}$ is given explicitly by
\begin{equation}
 G_{\mu\nu}^{GB} =  -\frac{1}{2}\mathcal{L} _{GB}g_{\mu\nu}+2\left[ -R_{\mu\rho\sigma\eta}{R_{\nu}}^{\rho\sigma\eta}-2R_{\mu\rho\nu\sigma}R^{\rho\sigma}-2R_{\mu\sigma}R_{\ \nu
}^{\sigma}+RR_{\mu\nu}\right],
\end{equation}
and $T_{\mu\nu}$ is the energy-momentum tensor of the matter that we consider as modified Chaplygin gas. We note that the divergence of EGB tensor $G_{\mu \nu}^{GB}$ vanishes. Here, we want to obtain 5D static spherically symmetric solutions of Eq.~(\ref{ee}) in the background of the modified Chaplygin gas and investigate the related properties. We assume that the metric has the form:
\begin{equation}
ds^2=-f(r)dt^2+\frac{1}{f(r)} dr^2+r^2 d\Omega^2_{3},\label{dsf}
\end{equation}
where $d\Omega_{3}$ is a line element on a $3$ dimensional hypersurface with constant scalar curvature $\kappa=1,0$ and $-1$, respectively for spherical, flat and hyperbolic spaces.  Using this metric ansatz, the EGB tensor $G_{\mu\nu}$ can be expressed as
\begin{eqnarray}\label{Gmunu}
&{G_t}^t={G_r}^r= \frac{3}{2r^2}\left[r f' + 2 (f-\kappa)\right] - \frac{6 \alpha}{r^3} \left[(f-\kappa)f'\right]+\Lambda,
\nonumber \\
&{G_{\theta}}^{\theta}={G_{\phi}}^{\phi} = {G_{\varphi}}^{\varphi} = \frac{1}{2r^2} \left[r^2f''+4rf'+2(f-\kappa)\right]-\frac{2\alpha}{r^2} \left[ (f-\kappa)f''+f'^2\right]+\Lambda.
\end{eqnarray}

\subsection{Modified Chaplygin gas surrounding a black hole}
We study the MCG with the equation of state (EoS) $p=A\rho-\frac{B}{\rho^\beta}$ \cite{Benaoum:2002zs,Debnath:2004cd}, where $A,B$ are positive parameters and $\beta$ stays in the interval $0\leq\beta\leq1$. For 5D spherically-symmetric spacetime, the energy-momentum tensor components of the MCG should have the general expression
\begin{equation}
{T}_t{}^t=\chi(r),~~{T}_t{}^i=0,~~{T}_i{}^j=\xi(r)\frac{r_ir^j}{r_nr^n}+\eta(r)\delta_i{}^j,\label{EMTMCG1}
\end{equation}
where the form for energy-momentum tensor was first considered by Kiselev when studying static spherically-symmetric quintessence surrounding a black hole~\cite{KiselevCQG2003}. Since we are considering static spherically-symmetric spacetime, the $r-r$ component of the energy-momentum tensor should be equal to the $t-t$ component, i.e.,
\begin{align}
{{T}_t}^t={{T}_r}^r=-\rho(r),
\label{EMTMCGtr}
\end{align}
If one takes isotropic average over the angles,
\begin{equation}
\langle r_i r^j\rangle=\frac{r_n r^n\delta_i{}^j}{4},\label{rAverage}
\end{equation}
one obtains
\begin{equation}
\langle {T}_i{}^j\rangle=\Big(\frac{\xi(r)}{4}+\eta(r)\Big)\delta_i{}^j=p(r)\delta_i{}^j
=\Big(A\rho(r)-\frac{B}{[\rho(r)]^\beta}\Big)\delta_i{}^j.\label{TMCGaverage}
\end{equation}
Considering Eqs.~(\ref{EMTMCGtr}) and (\ref{TMCGaverage}), $\xi(r)$ and $\eta(r)$ should be expressed as
\begin{equation}
\xi(r)=a_1\rho(r)+\frac{b_1}{[\rho(r)]^\beta},~~\eta(r)=a_2\rho(r)+\frac{b_2}{[\rho(r)]^\beta},\label{xieta}
\end{equation}
with the parameters $a_i$ and $b_i$, constrained by Eqs.~(\ref{EMTMCGtr}) and (\ref{TMCGaverage}), yielding:
\begin{equation}
a_1=-\frac{4}{3}(A+1),~~a_2=\frac{1+4A}{3},~~b_1=\frac{4}{3}B,~~b_2=-\frac{4}{3}B.
\end{equation}
Thus the angular components of the energy-momentum tensor are obtained as
\begin{align}
{{T}_{\theta}}^{\theta}={{T}_{\phi}}^{\phi}={{T}_{\varphi}}^{\varphi}=\frac{1+4A}{3}\rho(r)-\frac{4B}{3[\rho(r)]^\beta}.
\label{EMTMCGangular}
\end{align}

\subsection{Exact solutions}
Combining Eqs.~(\ref{Gmunu}), (\ref{EMTMCGtr}) and~(\ref{EMTMCGangular}), one obtains the EGB gravitational equations:
\begin{eqnarray}\label{EGBequations}
&\frac{3}{2r^2}\left[r f' + 2 (f-\kappa)\right] - \frac{6 \alpha}{r^3} \left[(f-\kappa)f'\right]+\Lambda=-\rho,~\nonumber\\
&\frac{1}{2r^2} \left[r^2f''+4rf'+2(f-\kappa)\right]-\frac{2\alpha}{r^2} \left[(f-\kappa)f''+f'^2\right]+\Lambda=\frac{1+4A}{3}\rho-\frac{4B}{3\rho^{\beta}}.
\end{eqnarray}
\begin{figure*}[htb]
\begin{tabular}{ c c c c}
\qquad\includegraphics[width=0.40\linewidth]{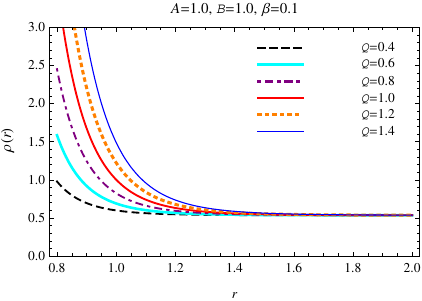}\quad
\includegraphics[width=0.40\linewidth]{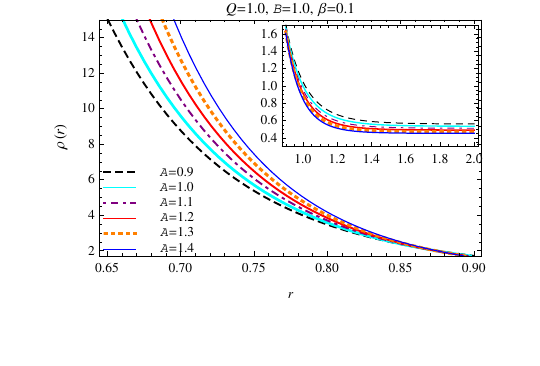}
\\
\qquad\includegraphics[width=0.40\linewidth]{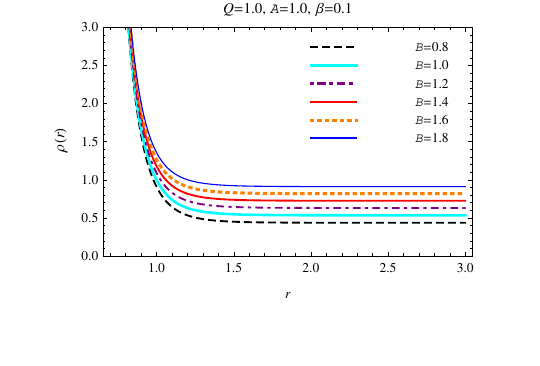}\quad
\includegraphics[width=0.40\linewidth]{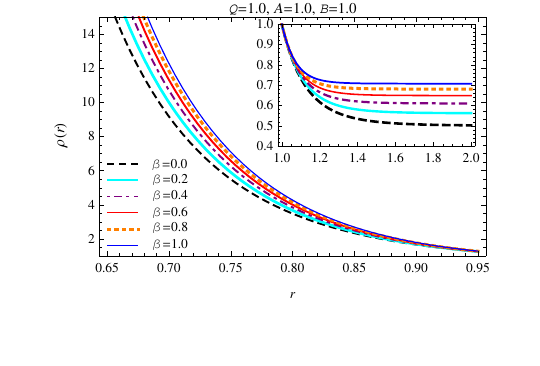}
\end{tabular}
\caption{\label{energydensity} Energy density of the MCG as function of the radial coordinate.}
\end{figure*}Thus, we have two unknown functions $f(r)$ and $\rho(r)$ which can be determined analytically by the above two differential equations. Now, by solving the set of differential equations~(\ref{EGBequations}), one first easily obtains the solution for the energy density of MCG:
\begin{equation}
\rho(r)=\left( \frac{1}{1+A}\left[B+\left(\frac{Q}{r^4}\right)^{(1+A)(1+\beta)}\right] \right)^{\frac{1}{1+\beta}}, \label{MCGenergydensity}
\end{equation}
where $Q>0$ is an integration constant. We observe that $\rho(r)\rightarrow \left(\frac{B}{1+A}\right)^{\frac{1}{1+\beta}}$ when $r\rightarrow\infty$, which means that the MCG acts like a cosmological constant very far from the black hole, and it gathers more densely as it moves toward the black hole because of the gravitation, as displayed in Fig.~\ref{energydensity}.

\begin{figure*}[htb]
\begin{tabular}{ c c c c}
\includegraphics[width=0.325\linewidth]{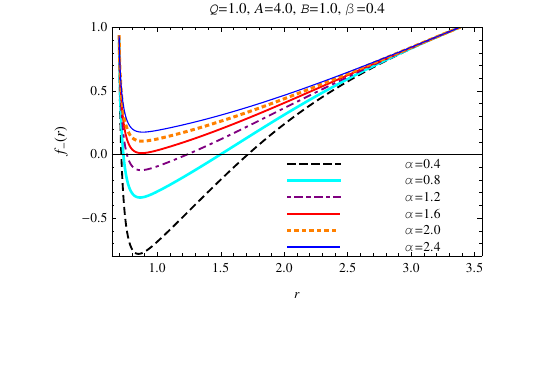}
\includegraphics[width=0.325\linewidth]{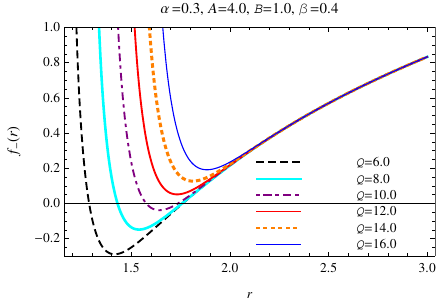}
\includegraphics[width=0.325\linewidth]{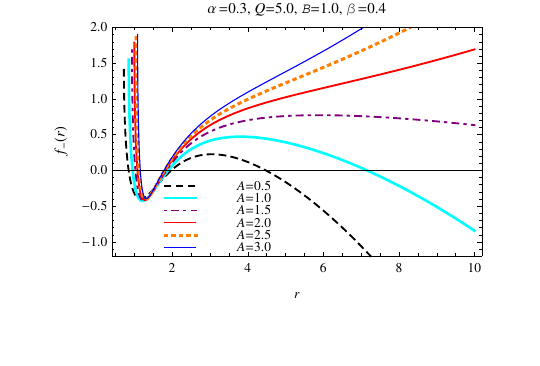}\quad
\\
\includegraphics[width=0.325\linewidth]{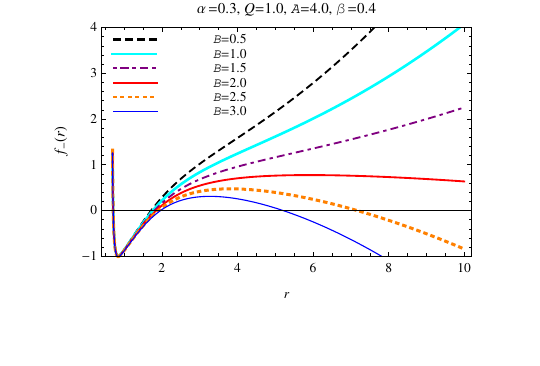}\quad
\includegraphics[width=0.325\linewidth]{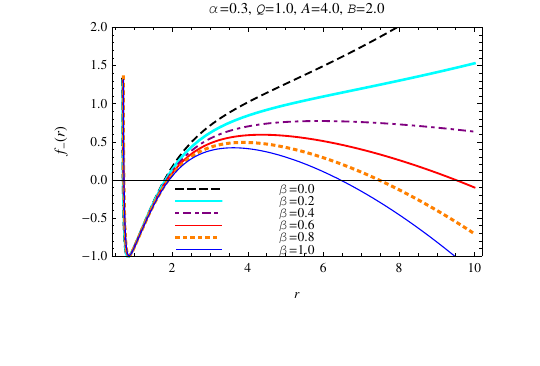}\quad
\end{tabular}
\caption{\label{frvarySABbeita}The $f_{-}(r)$ function in 5D Einstein-Gauss-Bonnet gravity with a background of modified Chaplygin gas for different values of $\alpha$, $Q$, $A$, $B$ and $\beta$. Here we set $m=2.0$, $\Lambda=-0.5$ and $\kappa=1$.}
\end{figure*}
\begin{figure*}[htb]
\begin{tabular}{ c c c}
\includegraphics[width=0.325\linewidth]{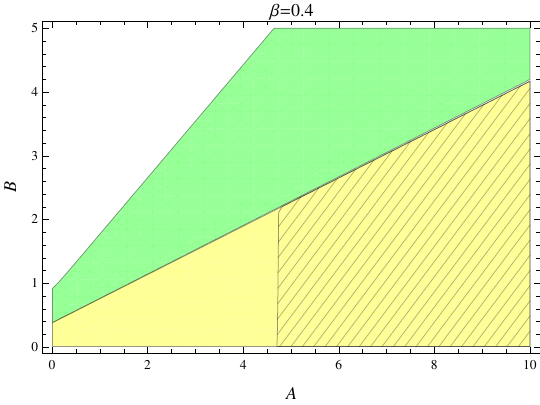}
\includegraphics[width=0.325\linewidth]{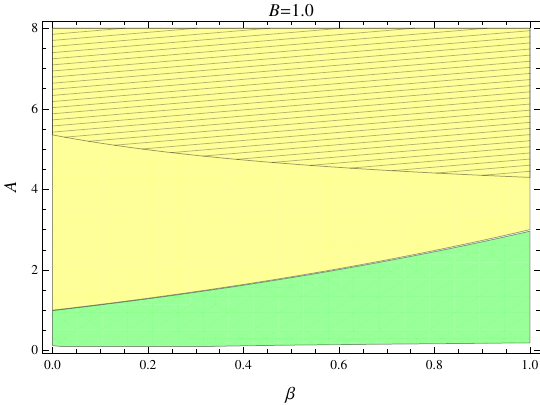}
\includegraphics[width=0.325\linewidth]{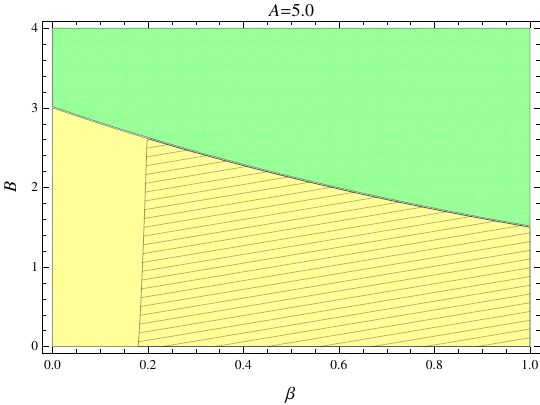}
\end{tabular}
\caption{\label{regionAdS}The MCG parametric regions for asymptotically de Sitter (green colored) and anti-de Sitter (yellow colored) black hole solution. The large yellow colored region correspond to $\kappa=1$ and $\kappa=0$, whereas the embedded meshed regions correspond to $\kappa=-1$; The green colored regions correspond to $\kappa=1$. Here we set $m=2.0$, $\alpha=0.2$, $Q=2.0$  and $\Lambda=-0.5$.}
\end{figure*}
\begin{figure*}[htb]
\begin{tabular}{ c c c}
\includegraphics[width=0.34\linewidth]{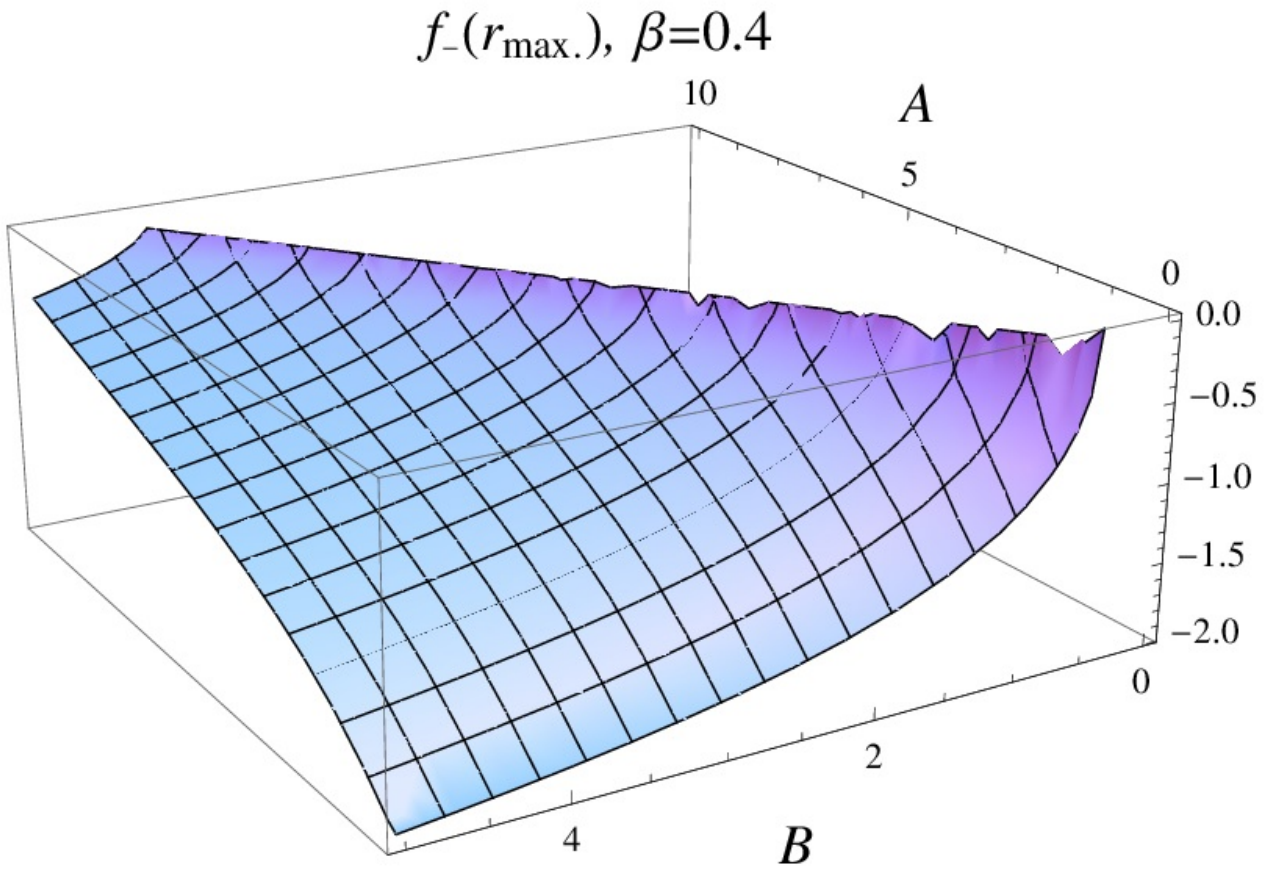}
\includegraphics[width=0.32\linewidth]{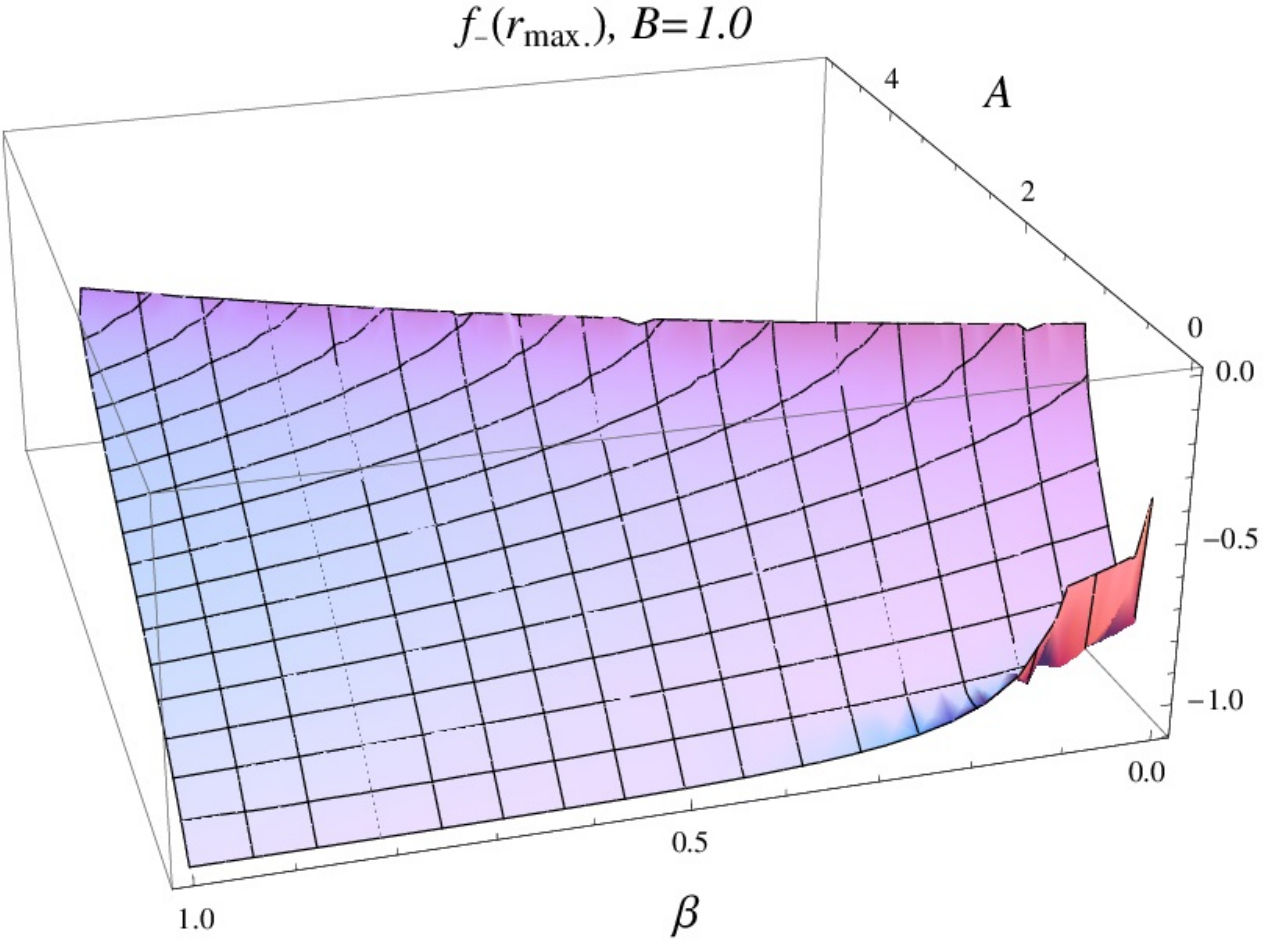}
\includegraphics[width=0.32\linewidth]{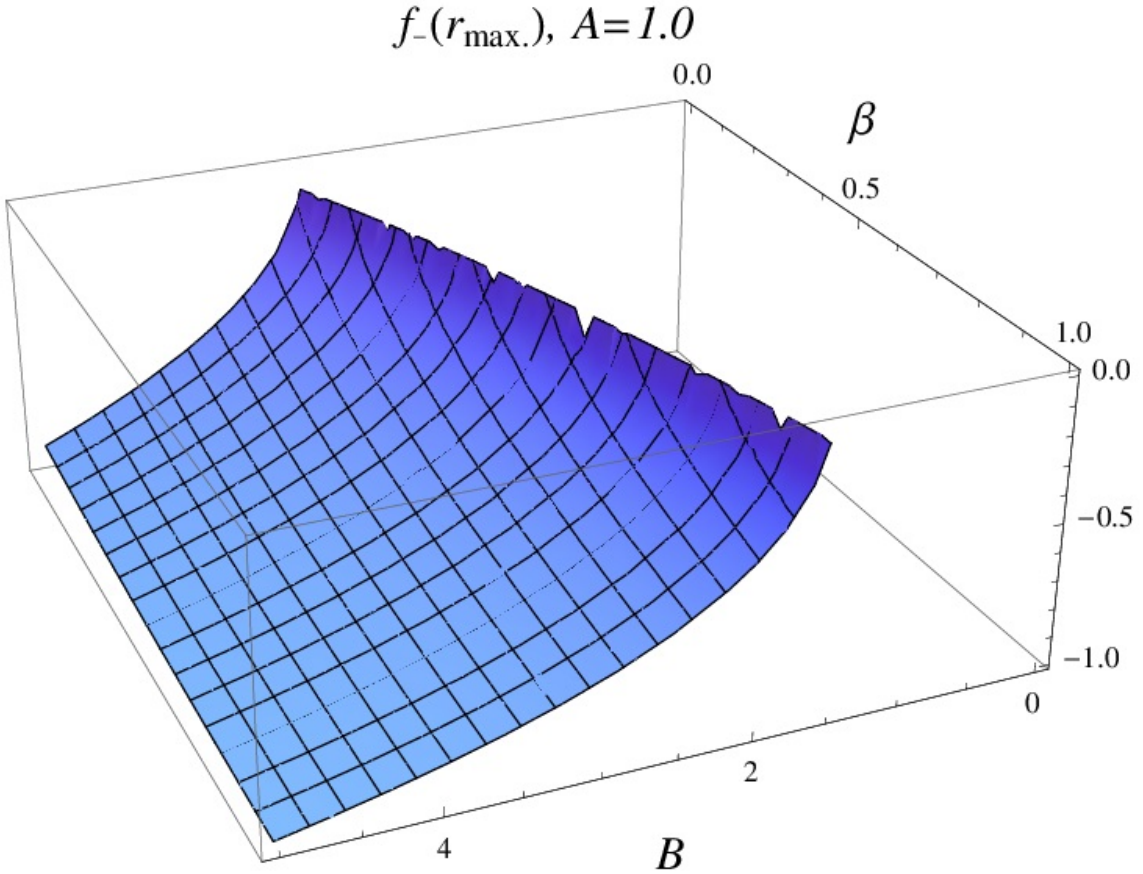}
\end{tabular}
\caption{\label{fredS}The values of $f_{-}(r)$ at the maximum point $r=r_{max.}$ for asymptotically de Sitter condition with $\kappa=0$. Here we set $m=2.0$, $\alpha=0.2$, $Q=2.0$ and $\Lambda=-0.5$.}
\end{figure*}

Substituting Eq.~(\ref{MCGenergydensity}) into the first differential equation in Eq.~(\ref{EGBequations}), we obtain two branches for the solution of $f(r)$:
\begin{equation} \label{twofbranches}
f_{\pm}(r) = \kappa+\frac{r^{2}}{4{\alpha}}\left(1\pm
\sqrt{1+\frac{16\alpha m}{r^{4}}+\frac{4\alpha}{3}\Lambda +\frac{4\alpha}{3}\Big(\frac{B}{1+A}\Big)^{\frac{1}{1+\beta}}\mathcal{F}}\right),
\end{equation}
where $\mathcal{F}$ represents the hypergeometric function and is defined as
\begin{eqnarray}\label{hypergeometric}
\mathcal{F}&=&_{2}F_{1}\left(\left[-\frac{1}{1+\beta},-\frac{1}{w}\right],1-\frac{1}{w},-\frac{1}{B}\left(\frac{Q}{r^4}\right)^{w}\right),~\nonumber\\
w&=&(1+A)(1+\beta).
\end{eqnarray}
We note that, $m$ is a parameter proportional to the mass of the black hole. The ADM mass $M$ of a 5D black hole relates to the mass parameter $m$ by $M=3\Sigma_3m$, where $\Sigma_3$ is the volume of the unit sphere in $\mathbb{R}^3$. To study the asymptotic behavior of $f_\pm(r)$, we take $r\rightarrow\infty$ and find that
\begin{equation}
f_{\pm}(r)\rightarrow \kappa+\frac{r^{2}}{4{\alpha}}\left(1\pm
\sqrt{1+\frac{4\alpha}{3}\Lambda+\frac{4\alpha}{3}\Big(\frac{B}{1+A}\Big)^{\frac{1}{1+\beta}}}\right).
\end{equation}
In the limit $\alpha \rightarrow 0$, the negative branch of the solution (\ref{twofbranches}) reduces to the 5D general relativity solution surrounded by the MCG. Thus we only concern $f_{-}(r)$ since it's well behaved. The EGB black holes surrounded by the MCG are characterized by their mass parameter $(m)$, the Gauss-Bonnet coupling constant ($\alpha$) and the four MCG parameters ($Q, A, B$ and $\beta$). The $f_{-}(r)$ solution with varying values of $\alpha$, $Q$, $A$, $B$ and $\beta$ is plotted in Fig.~\ref{frvarySABbeita}. We are inferred from the figures that, the black hole solution could be asymptotically anti-de Sitter or de Sitter according to the related parameters. It should be noted that, the radicand in Eq.~(\ref{twofbranches}) should be non-negative to keep the solution real-valued. For given mass, Gauss-Bonnet and MCG parameters, the radicand decreases to zero at the so-called branch singularity \cite{Torii:2005xu,Torii:2005nh}. We can conclude from Fig.~\ref{frvarySABbeita} that, the Gauss-Bonnet parameter $\alpha$ and the MCG parameter $Q$ significantly affect the near-field spacetime structure, thus affect the existence of black hole solution, while the MCG parameters $A$, $B$ and $\beta$ affect the asymptotic behavior of $f_{-}(r)$ at the far-field region, thus affect the type of black hole solution. $\alpha$ and $Q$ are liable to affect the positions of inner horizon and black hole horizon, while $A$, $B$ and $\beta$ are liable to affect the positions of black hole horizon and cosmological horizon.
We plot the MCG parameters regions for asymptotically (anti-)de Sitter black hole solutions in Fig.~\ref{regionAdS}. For asymptotically anti-de Sitter black hole solutions with $\kappa=1$ or $\kappa=0$, the MCG parameters should satisfy $\Lambda+\Big(\frac{B}{1+A}\Big)^{\frac{1}{1+\beta}}\leq0$ and $f_{-}(r_{min.})<0$, while for $\kappa=-1$, additional condition $f_{-}(r_{sing.})\geq0$ should be satisfied, where $f_{-}(r_{min.})$ and $f_{-}(r_{sing.})$ are respectively the values of $f_{-}(r)$ at the minimum point $r=r_{min.}$ and the branch singularity $r=r_{sing.}$. For asymptotically de Sitter black hole solutions with $\kappa=1$, the MCG parameters should satisfy $\Lambda+\Big(\frac{B}{1+A}\Big)^{\frac{1}{1+\beta}}>0$, $f_{-}(r_{min.})\leq0$ and $f_{-}(r_{max.})>0$, while for $\kappa=0$ and $\kappa=-1$, the value of $f_{-}(r)$ at the maximum point $r_{max.}$ keeps negative, thus there is no corresponding black hole solution, as shown in Fig.~\ref{fredS}.

\section{Thermodynamics}
\label{section3}
In this section, we discuss the thermodynamical properties of 5D MCG-surrounded black hole within Einstein-Gauss-Bonnet framework. It should be noted here that the MCG is always introduced in cosmology to illustrate the accelerating expansion of the Universe, however if we assume the MCG as a natural being under some mechanism, regardless of its cosmological implications, the spacetime could be asymptotically anti-de Sitter with the inclusion of a non-vanishing cosmological constant. In this work, we study the asymptotically anti-de Sitter case for mathematical completeness of the model. For asymptotically anti-de Sitter solutions, phase structure can be explained by critical quantities that are extracted in the extended phase space. We obtain critical values of pressure, volume, and temperature and investigate the effects of both the Gauss-Bonnet gravity and the MCG on these values. We plot $P-r_h$ and $G-T$ diagrams to study the phase transitions of these thermodynamical systems. For asymptotically de Sitter solutions, we study their thermodynamics for each horizon present in the spacetime as if they were independent thermodynamic systems characterized by their own temperature, and then show that these systems are not independent and various thermodynamic quantities in them are in fact entangled.

\subsection{Asymptotically anti-de Sitter solutions}
In order to investigate the phase structure of the solutions, one can employ the approach in which the cosmological constant is a thermodynamic variable corresponding to thermodynamical pressure with the relation
\begin{equation}
P=-\Lambda
\end{equation}
in the units $8\pi G_5 = c = 1$.

From Eq.~(\ref{twofbranches}), the mass of the black hole is obtained in terms of the horizon radius ($r_h$):
\begin{eqnarray}\label{M1}
m=\frac{ r_{h}^2 }{2}\left[\kappa+\frac{2\alpha\kappa^2}{r_{h}^2}+\frac{P}{6}r_{h}^2-\frac{r_{h}^2}{6}\Big(\frac{B}{1+A}\Big)^{\frac{1}{1+\beta}}\mathcal{F}(r_h) \right].
\end{eqnarray}
The Hawking temperature for the asymptotically anti-de Sitter black hole surrounded by the MCG can be calculated as
\begin{equation}\label{temp1}
T=\frac{r_h}{2\pi(r_h^{2}+4\alpha\kappa)}\left[\kappa+\frac{P}{3}r_h^2-\frac{r_h^2}{3}\rho(r_h)\right].
\end{equation}
Thus the pressure $P$ can be deduced as
\begin{equation}\label{pressure}
P=\frac{6\pi(r_h^{2}+4\alpha\kappa)}{r_h^{3}}T+\rho(r_h)-\frac{3\kappa}{r_h^{2}}.
\end{equation}
To compare it with the van der Waals equation, we make a series expansion for the van der Waals equation with the inverse of specific volume $v$,
\begin{equation}\label{vdWpressure}
P=\frac{T}{v-b}-\frac{a}{v^2}\approx\frac{T}{v}+\frac{bT}{v^2}-\frac{a}{v^2}+O(v^{-3}).
\end{equation}
By comparing Eq.~(\ref{pressure}) and (\ref{vdWpressure}), one can identify the specific volume $v$ with the horizon radius of the black holes as $v=\frac{r_h}{6\pi}$.

The critical point is determined as the inflection point in the $P-V$ diagram, i.e.,
\begin{equation}\label{criticalcondition}
\frac{\partial P}{\partial r_h}\Big|_{T=T_C,r_h=r_C}=\frac{\partial^{2}P}{\partial r_h^2}\Big|_{T=T_C,r_h=r_C}=0,
\end{equation}
where we have used the subscript `$C$' to stand for the quantities at the critical point. From Eq.~(\ref{pressure}), we can obtain the critical temperature
\begin{equation}\label{criticaltemp}
T_C=\frac{1}{\pi r_C (r_C^2+12\alpha \kappa)}\left[r_C^2\kappa -\frac{2}{3}Q\rho(r_C)^{-\beta}\left(\frac{Q}{r_C^4}\right)^{A+\beta+A\beta}\right],
\end{equation}
and the condition for the critical horizon radius $r_C$
\begin{eqnarray}\label{criticalradius}
&&8\beta K(r_C)^2 (r_C^2+12\alpha\kappa)-2\left[1+4\beta+4A(1+\beta)\right]K(r_C)(r_C^2+12\alpha\kappa)\rho(r_C)^{1+\beta} \nonumber\\
&&-4K(r_C)r_C^2\rho(r_C)^{1+\beta}+3\kappa\left(1-\frac{12\alpha\kappa}{r_C^2}\right)\rho(r_C)^{1+2\beta}=0,
\end{eqnarray}
with
\begin{equation}\label{Krc}
K(r_C)\equiv \left(\frac{Q}{r_C^4}\right)^{(1+A)(1+\beta)}.
\end{equation}

The entropy of the EGB black hole surrounded by the MCG, reads
\begin{eqnarray}\label{SEGB}
S=\frac{\Sigma_3r_h^3}{4\mathrm{G_5}}\left(1+\frac{12\alpha\kappa}{r_h^2}\right),
\end{eqnarray}
where $\Sigma_3r_h^3$ represents the horizon area of 5D black hole.

From Eq.~(\ref{pressure}), it is easy to see that for the Ricci flat black hole with $\kappa=0$, the pressure is a monotonic function of the horizon radius $r_h$, thus there does not exist any phase transition and therefore no critical point exists in this case. For hyperbolic topology with $\kappa=-1$, the non-negative definiteness of black hole entropy~(\ref{SEGB}) gives the constraint on the horizon radius,
\begin{equation}\label{Krc}
r_h^2\geq12\alpha.
\end{equation}
With this constraint, one can see that the critical temperature~(\ref{criticaltemp}) is always negative. Thus we conclude that in the case $\kappa=-1$, there does not exist any critical point and no phase transition happens. We can also obtain this conclusion by observing the derivative of~(\ref{pressure})
\begin{equation}\label{Dpressure}
\frac{\partial P}{\partial r_h}=-6\pi r_h^{-4}(r_h^{2}-12\alpha)T+\rho'(r_h)-6{r_h^{-3}},
\end{equation}
and noting the fact that when $\kappa=-1$, the pressure is a monotonic function of horizon radius.

To study the effects of different parameters on critical values, we present the numerical results about critical values of volume, temperature and pressure in Tables~\ref{varyingS}-\ref{varyingalpha}.
\begin{table}[!htbp]
    \centering
    \begin{tabular}{ccccc}
        \hline
        $Q$ & $v_C$ & $T_C$ & $P_C$ & $\frac{P_C v_C}{T_C}$ \\
        \hline
        0.500000 & 0.130103 & 0.064965 & 0.727810 & 1.457565 \\
        1.000000 & 0.130732 & 0.064922 & 0.727445 & 1.464835 \\
        1.500000 & 0.131896 & 0.064841 & 0.726750 & 1.478319\\
        2.000000 & 0.133516 & 0.064720 & 0.725738 & 1.497172 \\
        \hline
    \end{tabular}
    \caption{Critical quantities for varying $Q$. Here we set $A=1.0, B=1.0, \beta=0.2, \alpha=0.5$. }\label{varyingS}
\end{table}
\begin{table}[!htbp]
    \centering
    \begin{tabular}{ccccc}
        \hline
        $A$ & $v_C$ & $T_C$ & $P_C$ & $\frac{P_C v_C}{T_C}$ \\
        \hline
        0.600000 & 0.132884 & 0.064704 & 0.840395 & 1.725929 \\
        0.900000 & 0.131052 & 0.064895 & 0.751731 & 1.518078 \\
        1.200000 & 0.130335 & 0.064952 & 0.684846 & 1.374242 \\
        1.500000 & 0.130078 & 0.064968 & 0.632606 & 1.266590 \\
        \hline
    \end{tabular}
    \caption{Critical quantities for varying $A$. Here we set $Q=1.0, B=1.0, \beta=0.2, \alpha=0.5$. }\label{varyingA}
\end{table}
\begin{table}[!htbp]
    \centering
    \begin{tabular}{ccccc}
        \hline
        $B$ & $v_C$ & $T_C$ & $P_C$ & $\frac{P_C v_C}{T_C}$ \\
        \hline
        0.600000 & 0.130798 & 0.064918 & 0.532839 & 1.073582 \\
        0.900000 & 0.130745 & 0.064921 & 0.680263 & 1.369983 \\
        1.200000 & 0.130709 & 0.064924 & 0.819548 & 1.649975 \\
        1.500000 & 0.130683 & 0.064926 & 0.953080 & 1.918366 \\
        \hline
    \end{tabular}
    \caption{Critical quantities for varying $B$. Here we set $Q=1.0, A=1.0, \beta=0.2, \alpha=0.5$. }\label{varyingB}
\end{table}
\begin{table}[!htbp]
    \centering
    \begin{tabular}{ccccc}
        \hline
        $\beta$ & $v_C$ & $T_C$ & $P_C$ & $\frac{P_C v_C}{T_C}$ \\
        \hline
        0.200000 & 0.130732 & 0.064922 & 0.727445 & 1.464835 \\
        0.400000 & 0.130201 & 0.064961 & 0.776051 & 1.555438 \\
        0.600000 & 0.130025 & 0.064971 & 0.815055 & 1.631148 \\
        0.800000 & 0.129971 & 0.064974 & 0.847054 & 1.694415 \\
        \hline
    \end{tabular}
    \caption{Critical quantities for varying $\beta$. Here we set $Q=1.0, A=1.0, B=1.0, \alpha=0.5$. }\label{varyingbeta}
\end{table}
\begin{table}[!htbp]
    \centering
    \begin{tabular}{ccccc}
        \hline
        $\alpha$ & $v_C$ & $T_C$ & $P_C$ & $\frac{P_C v_C}{T_C}$ \\
        \hline
        0.200000 & 0.090666 & 0.100939 & 0.955067 & 0.857862 \\
        0.500000 & 0.130732 & 0.064922 & 0.727445 & 1.464835 \\
        0.800000 & 0.164546 & 0.051360 & 0.665350 & 2.131636 \\
        1.100000 & 0.192807 & 0.043804 & 0.636978 & 2.803697 \\
        \hline
    \end{tabular}
    \caption{Critical quantities for varying $\alpha$. Here we set $Q=1.0, A=1.0, B=1.0, \beta=0.2$. }\label{varyingalpha}
\end{table}
\begin{figure*}
\begin{tabular}{ c c}
\includegraphics[width=0.49\linewidth]{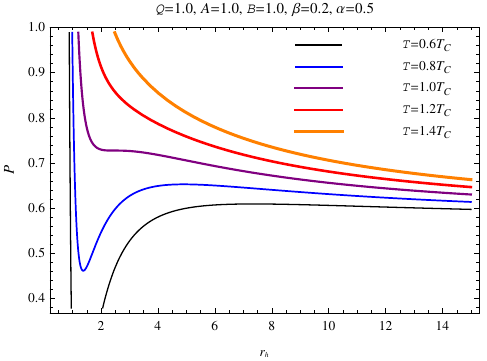}
\includegraphics[width=0.49\linewidth]{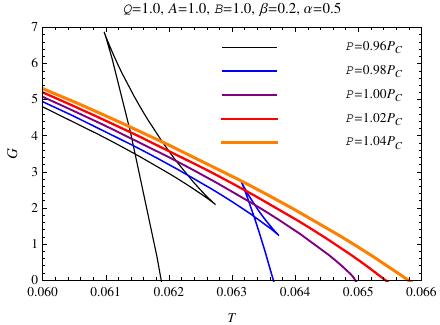}
\end{tabular}
\caption{\label{phasediagram}$P-r_h$ (left) and $G-T$ (right) diagrams for 5D asymptotically anti-de Sitter EGB black holes surrounded by MCG.}
\end{figure*}

The Gibbs free energy reads
\begin{eqnarray}\label{GibbsEGB}
G=G(T,P,A,B,\beta,\alpha)=H-TS=3\Sigma_{3}m-TS.
\end{eqnarray}

The corresponding $P-r_h$ diagram drawn in the left plot of Fig.~\ref{phasediagram} is exactly the same as the $P-V$ diagram of the van der Waals liquid-gas system. For black hole with fixed temperature lower than the critical one, there are two sections whose pressure decreases as the horizon radius increases, one is in the small radius region, corresponding to fluid phase in the van der Waals system, and the other is in the large radius region, corresponding to the gas phase. Such two sections have positive compression coefficients, thus representing stable phases. Between these two stable phases, there is an unstable phase with a negative compression coefficient. For appropriate values of pressure, the isothermal line allows two physical horizon radii. Therefore, the so-called small black hole/large black hole phase transition occurs. Such phase transition is first order for $T < T_C$, while it becomes second order at $T_C$ just as the same as the case in the van der Waals system. Above the critical temperature $T_C$, the black holes are always in the gas phase and no phase transition happens. We also plot the Gibbs free energy defined in Eq.~(\ref{GibbsEGB}) in the right plot of Fig.~\ref{phasediagram} as a function of temperature for different pressures. We observe from the $G-T$ plot that the ``swallow tail'' appears for $P<P_C$, which is a typical feature of the first order phase transition, and at the critical point $P=P_C$, the ``swallow tail'' disappears.

\subsection{Asymptotically de Sitter solutions}
According to our analysis in section~\ref{section2}, asymptotically MCG-surrounded-EGB black hole solutions exist only for $\kappa=1$. A typical asymptotically de Sitter black hole has three horizons, an inner event horizon $r_i$, a black hole horizon $r_b$ and a cosmological horizon $r_c$. The existence of a cosmological horizon in addition to the black hole horizon means that the system associated with an observer living between the horizons is in a non-equilibrium state, thus an observer would find himself in a thermodynamic system characterized by two temperatures. The thermodynamic quantities associated with the three horizons are given by
\begin{equation}\label{temp1}
T_b=\frac{r_b}{2\pi(r_b^{2}+4\alpha)}\left[1+\frac{P}{3}r_b^2-\frac{r_b^2}{3}\rho(r_b)\right], S_b=\frac{\Sigma_3r_b^3}{4\mathrm{G_5}}\left(1+\frac{12\alpha}{r_b^2}\right),
\end{equation}
\begin{equation}\label{temp1}
T_c=\frac{-r_c}{2\pi(r_c^{2}+4\alpha)}\left[1+\frac{P}{3}r_c^2-\frac{r_c^2}{3}\rho(r_c)\right], S_c=\frac{\Sigma_3r_c^3}{4\mathrm{G_5}}\left(1+\frac{12\alpha}{r_c^2}\right),
\end{equation}
\begin{equation}\label{temp1}
T_i=\frac{-r_i}{2\pi(r_i^{2}+4\alpha)}\left[1+\frac{P}{3}r_i^2-\frac{r_i^2}{3}\rho(r_i)\right], S_i=\frac{\Sigma_3r_i^3}{4\mathrm{G_5}}\left(1+\frac{12\alpha}{r_i^2}\right),
\end{equation}
where the minus signs in $T_c$ and $T_i$ guarantee that the cosmological and inner horizon temperatures are nonnegative. The quantities analogous to the Gibbs free energy are given by
\begin{equation}\label{temp1}
G_b=M-T_bS_b,
\end{equation}
\begin{equation}\label{temp1}
G_c=-M-T_cS_c,
\end{equation}
\begin{equation}\label{temp1}
G_i=-M-T_iS_i.
\end{equation}
where $M$ is treated as the gravitational enthalpy of the black hole horizon, whereas $-M$ is treated as the gravitational enthalpy of the cosmological and inner horizons, considering that both $T_cdS_c$ and $T_idS_i$, which should appear in the thermodynamic first laws of cosmological and inner black hole horizons, have an opposite sign to what is customary in order to treat $M$ as the enthalpy \cite{KubiznakCQG2016}.

\begin{figure*}
\begin{tabular}{ c c c}
\includegraphics[width=0.32\linewidth]{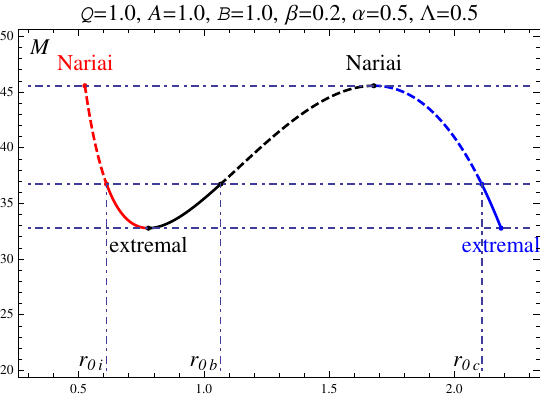}
\includegraphics[width=0.32\linewidth]{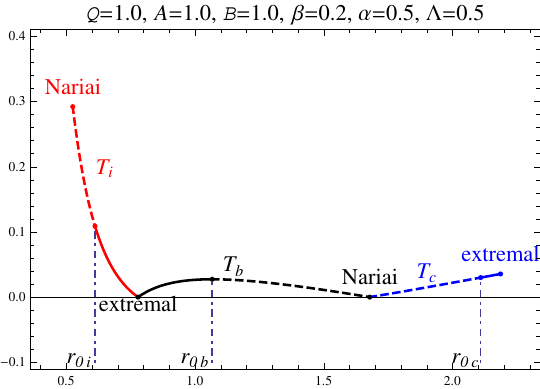}
\includegraphics[width=0.32\linewidth]{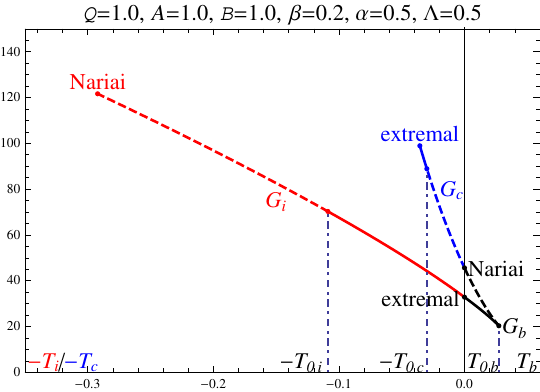}
\end{tabular}
\caption{\label{thermodeSitter}Thermodynamics of asymptotically de Sitter MCG-surrounded-EGB black hole. The black hole mass (\emph{Left}) and horizon temperatures (\emph{Middle}) are displayed as functions of corresponding horizons: inner (red), cosmological (blue), and black hole (black); dashed curves correspond to regions that are thermodynamically unstable. The cusp occurs for the black hole horizon $r_{0b}$; $r_{0i}$ and $r_{0c}$ are the corresponding inner and cosmological horizons. The corresponding Gibbs free energies are depicted as functions of horizon temperatures (\emph{Right}), displaying the thermodynamic behavior of all three horizons. The cusp occurs for the black hole horizon temperature $T_{0b}$; $T_{0i}$ and $T_{0c}$ are the corresponding temperatures of the inner and cosmological horizons.}
\end{figure*}

We note that the three horizons are in fact entangled, observing from Fig.~\ref{thermodeSitter} that when black hole mass decreases, the black hole horizon decreases, meanwhile the inner horizon and cosmological horizon increase. Thus the thermodynamics of asymptotically de Sitter MCG-surrounded-EGB black hole can be characterized by alone the black hole horizon. Several special cases of asymptotically de Sitter MCG-surrounded-EGB black hole are of importance. The extremal black hole limit corresponds to the case where the inner horizon coincides with the black hole horizon, $r_b=r_i$. It can be easily shown that in this case we have $T_b=T_i=0$. The second important case is the so called Nariai limit in which $r_b=r_c$ and consequently, $T_b=T_c=0$. The cusp of the Gibbs free energy curve corresponds to the maximum point of the black hole horizon temperature curve. There is no critical point for Gibbs free energy, thus there is no phase transition for asymptotically de Sitter MCG-surrounded-EGB black hole.

\section{Conclusion}
\label{section4}
In this study we have obtained an exact black hole solution that integrates the surrounding modified Chaplygin gas into the framework of 5D Einstein-Gauss-Bonnet theory. It is found that the spacetime structure could be asymptotically anti-de Sitter or de-Sitter according to the specific values of EGB and MCG parameters. Asymptotically anti-de Sitter black hole solutions exist for all cases of $\kappa=1,0$ and $-1$, while asymptotically de Sitter black hole solutions exist only for Ricci spherical ($\kappa=1$) case. The parametric regions for both kinds of black hole solutions were showed in Fig.~\ref{regionAdS}. We have studied the thermodynamic behavior of asymptotically anti-de Sitter black hole in the extended space by considering the cosmological constant as thermodynamical pressure and found that $P-V$ criticality and the small black hole/large black hole phase transition appear. We also herein studied the dependence of the critical exponents on the Gauss-Bonnet and MCG parameters. A typical asymptotically de Sitter black hole has three horizons, inner horizon, black horizon and cosmological horizon. We studied the thermodynamic properties of the three horizons as if they were independent thermodynamic systems and showed that these system were in fact entangled and could be captured by a single Gibbs free energy $G$. By examining the $G-T$ relation, we found that there was no critical point thus no phase transition for $5D$ asymptotically de Sitter black hole.

\appendix
\section{Einstein-Gauss-Bonnet black holes in \emph{D}-dimensional spacetime}
\label{appendixA}
The energy density of the modified Chaplygin gas shows
\begin{equation}
\rho(r)=\left[\frac{1}{1+A}\left(B+\left[\frac{Q}{r^{D-1}}\right]^{(1+A)(1+\beta)}\right)\right]^{\frac{1}{1+\beta}}. \label{dfdensity}
\end{equation}

The solution for the metric function is obtained as
\begin{equation} \label{newtwofbranches}
f_{\pm}(r) = \kappa+\frac{r^{2}}{2{\alpha}(D-3)(D-4)}\left(1\pm
\sqrt{\mathcal{H}}\right)
\end{equation}
with
\begin{eqnarray}
\mathcal{H}&=&1+\frac{4\alpha(D-3)(D-4)}{D-2}\left[\frac{2 M}{\Sigma_{D-2}r^{D-1}}+\frac{2\Lambda}{D-1} +\frac{2}{D-1}\Big(\frac{B}{1+A}\Big)^{\frac{1}{1+\beta}}\mathcal{F}\right],~\nonumber\\
\mathcal{F}&=&_{2}F_{1}\left(\left[-\frac{1}{1+\beta},-\frac{1}{w}\right],1-\frac{1}{w},-\frac{1}{B}\left(\frac{Q}{r^{D-1}}\right)^{w}\right),~\nonumber\\
w&=&(1+A)(1+\beta),
\end{eqnarray}
where $\Sigma_{D-2}$ denotes the volume of the unit $(D-2)$ sphere. Again, only the minus branch solution $f_{-}(r)$ is considered as a black hole solution.

The black hole horizon $r_h$ for asymptotically anti-de Sitter solutions satisfies $f_-(r_h)=0$, while the black hole horizon $r_b$, cosmological horizon $r_c$ and inner horizon $r_i$ for asymptotically de Sitter solutions satisfy the equation $f_-(r_{b,c,i})=0$. The ADM mass of the black hole in terms of the several horizons reads
\begin{equation}
\begin{aligned}
M&=\frac{\Sigma_{D-2}}{2}\left[(D-2)r_{h,b,c,i}^{D-3}\kappa+(D-2)(D-3)(D-4)r_{h,b,c,i}^{D-5}\alpha\kappa^2-\frac{2\Lambda}{D-1}r_{h,b,c,i}^{D-1} \right.\\
&\left.-\frac{2r_{h,b,c,i}^{D-1}}{D-1}\left(\frac{B}{1+A}\right)^{\frac{1}{1+\alpha}}\mathcal{F}(r_{h,b,c,i})\right].\label{DdimensionalM}
\end{aligned}
\end{equation}
According to Eq.~(\ref{DdimensionalM}), the three horizons of asymptotically de Sitter black hole are entangled. The Hawking temperatures associated with different horizons are calculated as
\begin{equation}
T_{h,b}=\frac{r_{h,b}}{4\pi \mathcal{N}(r_{h,b})}\left[(D-5)\kappa+(D-3)(D-4)(D-5)\frac{\alpha\kappa^2}{r_{h,b}^2}+2\kappa-\frac{2}{D-2}r_{h,b}^2(\Lambda+\rho(r_{h,b}))\right],\label{T1}
\end{equation}
\begin{equation}
T_{c,i}=\frac{-r_{c,i}}{4\pi \mathcal{N}(r_{c,i})}\left[(D-5)\kappa+(D-3)(D-4)(D-5)\frac{\alpha\kappa^2}{r_{c,i}^2}+2\kappa-\frac{2}{D-2}r_{c,i}^2(\Lambda+\rho(r_{c,i}))\right],\label{T1}
\end{equation}
with $$\mathcal{N}(r_{h,b,c,i})=r_{h,b,c,i}^2+2(D-3)(D-4)\alpha\kappa.$$

The entropies are given by
\begin{equation}
S_{h,b,c,i}=2\pi(D-2)\Sigma_{D-2}r_{h,b,c,i}^{D-4}\left[\frac{r_{h,b,c,i}^2}{D-2}+2(D-3)\alpha\kappa\right].
\end{equation}

The Gibbs free energies related to different horizons are identified as
\begin{equation}\label{Ddimensionaltemp1}
G_{h,b}=M-T_{h,b}S_{h,b},
\end{equation}
\begin{equation}\label{Ddimensionaltemp2}
G_{c,i}=-M-T_{c,i}S_{c,i}.
\end{equation}

\section*{Acknowledgments}
This work is partly supported by the Special Foundation for Theoretical Physics Research Program of China (Grant No. 11847065), the Natural Science Foundation of Shanxi Province (No. 201901D211110).

\section*{Data Availability Statement}
No Data associated in the manuscript.

\end{document}